\documentclass{mem}
\usepackage{natbib}\usepackage{txfonts}\usepackage{balance}
\usepackage{graphicx}
\usepackage[a4paper]{hyperref}
\idline{75}{382}
\begin{document}
\def\teff{$T\rm_{eff }$}
\def\kms{$\mathrm {km s}^{-1}$}

\title{Accretion disc coronae in black hole binaries}

   \subtitle{}

\author{
J. \,Malzac\inst{1} 
          }

  \offprints{J. Malzac }

\institute{
Centre d'Etude Spatiale des Rayonnements, OMP, UPS, CNRS;  9 Avenue du Colonel Roche, 31028 Toulouse, France
\email{malzac@cesr.fr}
}

\authorrunning{Malzac }

\titlerunning{Accretion discs coronae}

\abstract{
Most of the luminosity of accreting black hole is emitted in the X-ray band. This radiation is believed to emerge, through inverse Compton process, from a hot  (Te $\sim$ 10$^8$ -10$^9$ K)  optically thin ($\tau_T \sim 1$)  plasma probably located in the immediate vicinity of the black hole. The mechanisms at work in this so called  Compton corona  can be unveiled through hard X-ray observations which have revealed a rich phenomenology. Depending on luminosity different spectral states are observed suggesting that the nature and geometry of the corona depends on mass accretion rate.  In many instance  the spectral behaviour as a function of luminosity show some degree of hysteresis. The mecanisms triggering  the transition between spectral states is very unclear although it could  be related to an evaporation/condensation equilibrium in an accretion disc corona system. From the observation of correlation between the X-ray and radio band , it appears that the Compton corona is intimately related to the formation of compact jets and probably constitutes the base of the jet.
}

\maketitle{}

\section{Spectral states and the structure of the accretion flow}

Accreting black holes are observed in two main spectral states (see e.g. Zdziarski \& Girelinski 2004). At luminosities exceeding a few percent of Eddington ($L_{Edd}$), the spectrum is dominated by a thermal component peaking at a few keV which is believed to be the signature of a geometrically thin optically thick disc (Shakura \& Sunyaev 1974).
At higher energies the spectrum is non-thermal and usually present a weak steep power-law component (photon index $\Gamma \sim 2.3-3$) extending at least to MeV energies, without any hint for a high energy cut-off. This soft power law is generally interpreted as inverse Compton up-scaterring of soft photons (UV, soft X)   by a non-thermal distribution of electrons. Since in this state the source is bright in soft X-rays and soft in hard X-rays it is called the High Soft State (hereafter HSS).

At lower luminosities (L$<$ 0.01 $L_{Edd}$) , the appearance of the accretion flow is very different: the spectrum can be modelled as a hard power-law $\Gamma \sim 1.5-1.9$
  with a cut-off at $\sim100$ keV. The $\nu F_{nu}$ spectrum then peaks around a hundred keV. 
 Since  the soft X-ray luminosity is faint and the spectrum is hard, this state is called the Low Hard State (hereafter LHS). LHS spectra are generally very well fitted by Comptonisation  (i.e. multiple Compton up-scatering) of soft  photons 
 by a Maxwellian distribution of  electrons in a hot ($kT_{e}\sim$ 100 keV)  plasma of Thomson optical depth $\tau_{T}$ of order unity. As it will be discussed below the LHS is asssociated with the presence of a compact radio jet that seem to be absent in the HSS. In addition to the dominant comptonisation spectrum there are other less 
prominent spectral features: a weak soft component associated to the thermal emission of a geometrically thin optically thick disk is occasionally detected below 1 keV as observed for instance  in Cygnus X-1 (Balucinska-Church et al. 1995) or in XTE J1118+480 (McClintock et al. 2001, Chaty et al. 2003). 
Finally in both states, reflection features are generally detected in the form of a Fe K$\alpha$ line peaking at 6.4 kev and a Compton reflection bump peaking at $\sim$ 30 keV. These components are believed to be produced when the hard X-ray emission of the corona interacts and is reflected by the cold thick accretion disc. They often appears to be broadened through special and general relativistic effects, in which case their origin must be very close to the black hole. These reflection features appear to be weaker in the LHS than in the HSS.

To sumarize, at high luminosities the accretion flow is in the HSS caracterized by a strong thermal disc and reflection component and a weak non-thermal (or hybrid thermal/non-thermal) comptonising corona. At lower luminosity the disc blackbody and reflection features are weaker,  while the corona is dominant and emits through thermal Comptonisation.
Beside the LHS and HSS, there are several other spectral states that often appear, but not always, when the source is about to switch from one of the two main states to the other. Those states are more complex and difficult to define. We refer the reader to McClintock \& Remillard (2006) and Belloni et al. (2005) for two different  spectral classifications based on X-ray temporal as well as spectral criteria and radio emission (Fender 2006).  In general their spectral properties  are intermediate between those of the  LHS and HSS.


The different spectral states are usually understood in terms of changes in the geometry of the accretion flow. The standard picture is that in the HSS there is a standard geometrically thin disc extending down to the last stable orbit and responsible for the dominant thermal emission. This disc is the source of soft seed photons for Comptonisation in small active coronal  regions located above an below the disc. Through magnetic buoyancy the magnetic field lines rise above the accretion disc, transporting a significant fraction of the accretion power into the corona where it is then dissipated in through magnetic reconnection (Galeev et al. 1977).  In the corona particles are accelerated.  A population of  high energy electrons is formed which then cool down by up scattering the soft photons coming from the disc. This produces the high energy non-thermal emission which in turn illuminates the disc forming strong reflection features (see e.g. Zdziarski \& Gierli{\'n}ski 2004)

In the LHS, the standard geometrically thin disc does not extend to the last stable orbit, instead, the weakness of the thermal features suggest that it is truncated at distances ranging from a few hundreds to a few thousands gravitational radii from the black hole (typically  1000--10000 km). In these inner parts the accretion flow takes the form of a hot geometrically thick, optically thin disc. A solution of such hot accretion flow was first discovered by Shapiro, Lightman and  Eardley (1976). In these hot acccretion flow the gravitational energy is converted in the process of viscous dissipation into the thermal energy of ions. The main coupling between the electrons and the ions is Coulomb collision which is rather weak in the hot thin plasma. Since radiative cooling of the ions is much longer than that of the electrons, the ions temperature is much higher than the electron temperature.  This two temperature plasma solution  is  thermally unstable (Pringle 1976)  but can be stabilized if advection of the hot gas into the black hole, or alternatively into an outflow, dominates the energy transfert for ions (Ichimaru 1977; Rees et al. 1982; Narayan \& Yi 1994, Abramowicz et al. 1995; Blandford \& Begelman 1999; Yuan 2001, 2003). In these different models of advection dominated accretion flows (ADAF) most of the power is not radiated, they are therefore radiatively  inefficents.  
The electron have a thermal distribution and cool down by Comptonisation of  the soft photons coming from the external geometrically thin disc, and IR-optical photons internally generated through self-absorbed synchrotorn radiation.  The balance between heating and cooling determines the electron temperature which are found to be of order of 10$^9$ K  as required to fit the spectrum.
The weak reflection features of the LHS are produced through illumination of the cold outer disc by the central source.  
By extension, the hot accretion flow of the LHS  is frequently designed as  "corona" despite the lack of a direct physical analogy with the  rarefied gaseous envelope of the sun and other stars.  

\begin{figure*}[t]
\resizebox{\hsize}{!}{\includegraphics[clip=true]{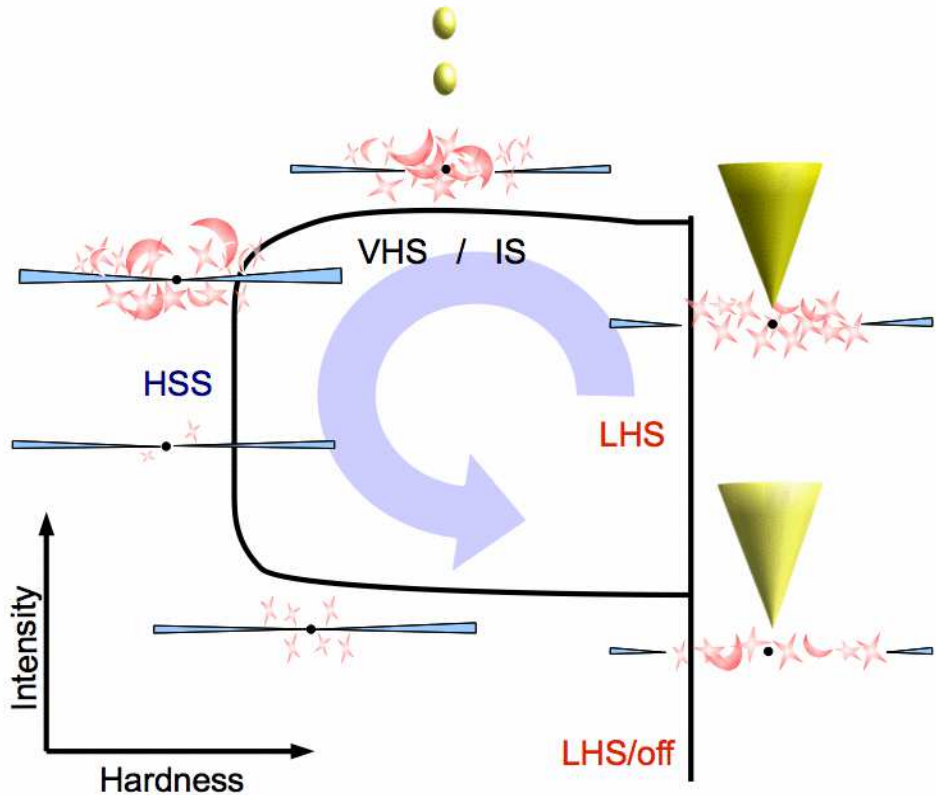}}
\caption{
\footnotesize
The black curve show the typical path followed by  a black hole X-ray transient in the hardness vs intensity diagram during an outburst. The various sketches illustrate the standard scenario for the evolution of the geometry of the corona-disc-jet system along this path (see text).}
\label{fig:hyste}
\end{figure*}

\section{An alternative model for the LHS: accretion disc corona}

The presence of the hot accretion disc is not the only possibility to explain the LHS. 
 It was suggested long ago that instead, a real accretion disc corona system similar to that of the HSS may reproduce the spectra as well (Bisnovatyi-Kogan \& Blinikov 1976; Liang \& Price 1977) . This would imply a cold geometrically thin disc extending down very close to the black hole in the LHS. Unlike in the HSS, this disc does not produce a strong thermal compoonent in the X-ray pectrum because it is too cold. It is cold because most of the accretion power is not dissipated in the disc, rather it is  transported away to power a strong corona and the compact jet. 
This, at least in principle, could be achieved either through transport via buyonancy of the magnetic field (Miller \& Stone 2000; Merloni \& Fabian 2002) or through the torque exerted on the accretion disc by a strong magnetic field threading the disc and driving the jet (Ferreira  1997). 

Due to the complexity of the accretion disc corona physics it is not possible to obtain simple analytical solutions giving the main properties of the corona unless  some parametrization of the energy transfert between the disc and the corona is used (see e.g.  Svensson \& Zdziarski 1994).
However, from an observer perspective it does not really matter because the appearence of the accretion disc corona does not depend on the details of the energy transport and dissipation mechanisms. Indeed, Haardt \& Maraschi (1991) realised  the existence of a strong radiative feedback between the cold disc and the hot corona. A fixed fraction of the power dissipated in the corona in the form of hard X-ray photons,  impinge on the cold disc where it is absorbed, heating the cold disc, and finally reemited in the form of thermal soft photons. A fraction of those soft photon re-enters the corona providing the major cooling effect to the coronna trough inverse Compton. As a consequence the cooling rate of the corona  scales like the heating rate. The coronal temperature is determined only by the geometry of the accretion disc corona (that control the fraction of coronal power that returns to the corona in the form of soft photons) . For instance an extended corona sandwiching the cold accretion disc, would intercept more cooling  photons from the accretion disc than a patchy corona made of a few compact  active regions covering a small fraction of the disc, and therefore would have a lower temperature and hence a harder comptonised X-ray spectra. 
During the nineties several group (Haardt \& Maraschi 1993; Stern et al. 1995 ; Poutanen et al. 1996)
 performed detailed computations of the resulting equilibrium spectra for various geometries. 
They concluded that the corona has to be patchy in order to produce spectra that hare hard enough (Haardt et al. 1994; Stern et al. 1995) even when the effects of disc ionisation are accounted for (Malzac et al 2005).
Yet, those accretion disc corona models had an important problem:  the production of an unobserved  strong thermal component due to reprocessing of the radiation illuminating the disc. In the case of an active region emitting isotropically, about half of the luminosity intercepts the disc and the thermal reprocessing component is comparable in luminosity to the primary emission, while in LHS spectra the thermal comonent from the disc is barely detectable. Morevover, the amplitude of the reflection features observed in the LHS are lower  than what is expected from an istotropic corona by at least a factor of 3, in some cases they are so weak that they are not even detected. 
This led to consider models where the coronal emission is not isotropic. Beloborodov (1999) suggested that the corona is unlikely to stay at rest with respect to the accretion disc. Due to the anisotropy of the dissipation process or simply to radiation pressure from the disc, the hot plasma is likely to be moving at midly relativistic velocities. Then, due to Doppler effects, the X-ray emission is strongly beamed in the direction of the plasma velocity. In the case of a velocity directed away from the accretion disc, (outflowing corona) , the reprocessing features (both reflection and thermalised radiation)  are strongly suppressed. Moreover, due to the reduced feedback from the disc, the corona is hotter, and harder spectra can be produced. 
Malzac, Beloborodov \& Poutanen (2001) performed detailled non-linear Monte-Carlo simulation of this dynamic accretion disc corona equilibrium, and compared the results with the observations. They found that compact actives regions of aspect ratio of order unity,  outflowing with a velocity of 30 percent of the speed of light could reproduce the LHS spectrum of Cygnus X-1. Moreover since the velocity of the coronal plasma controls both the strength of the reflection features and the feedback of soft cooling photons from the disc, it predicts a correlation between the slope of the the hard X-ray spectrum and  the amplitude of the reflection component. 
Such a correlation is indeed observed in several sources (Zdziarski et al. 1999, 2003) and is well matched by this model.

Recently the accretion disc corona models for the LHS obtained more observationnal  support, with the discovery of  relativistically broadened iron line in the LHS of GX339-4 (Miller et al., 2006). Such relativistically broadened lines require that disc illumination is taking place very close to the black hole. This observation suggests that, at least in some cases, a thin disc is present at or close to the last stable orbit in the LHS.

\section{The jet coronna connection}

Recent multi-wavelength observations
of accreting  black holes in the LHS have shown the presence of
an ubiquitous flat-spectrum radio emission (see e.g Fender 2006), that may
extend up to infrared and optical wavelengths. 
The properties the radio emission indicate it is likely  produced
by synchrotron emission from relativistic electrons in compact,
self-absorbed jets (Blandford \& Konigl, 1979; Hjellming \& Johnston 1988). 
This idea was confirmed by the discovery 
of a continuous and steady milliarcsecond 
compact jet around Cygnus X-1  (Sirling 2001).
Moreover, in LHS sources a tight 
correlation has been found between the hard X-ray and radio luminosities, holding over more than three decades in luminosity 
(Corbel et al. 2003; Gallo, Fender \& Pooley 2003).
In contrast, during HSS episodes the sources appear to be
radio weak (Tananbaum et al. 1972; Fender et al. 1999; Corbel 2000), 
suggesting that the Comptonising medium of the low/hard 
state is closely linked to the continuous ejection of matter in 
the form of a small scale jet.

When the importance of the connection between radio and X-ray emission was realised.
It was proposed that the hard X-ray emission could be in fact synchrotron emission in the jet, rather  than  comptonisation in a hot accretion flow/corona (Markoff, Falke, Fender 2001; Markoff et al. 2003). This model is able to explain  quantitatively the correlation between the X-ray and radio emission, it is also able to reproduce at least roughly the shape of the LHS X-ray spectrum of several sources.  However, it seems that synchrotron emission alone is not enough to reproduce the details of the X-ray spectra. In the most recent version of this model a thermal Comptonisation component was added which appears to provide a dominant contribution to the hard X-ray spectrum (Markoff et al., 2005).  This component is supposedly formed in the base of the jet which forms a hot plasma that present strong similarities with an accretion disc corona. 
In the context of accretion disc coronae/hot disc models the correlation between X-ray and radio emission, tells us that the corona and the compact jet of the LHS are intimately connected. 
A strong corona may be necessary to launch  a jet and/or could be  the physical location where the jet is accelerated or launched (Merloni \& Fabian 2002).  
Koerding et al. (2006) show that the observed X-radio correlation can be reproduced provided that the 
hot flow/corona is radiatively inefficient (i.e. luminosity scales like the square of mass accretion rate) and that a constant fraction of the accretion power goes into the jet (i.e. jet power scales like mass accretion rate). 
However there is presently no detailed model to explain how the physical connection between jet and corona works.  A possible basic explanation was proposed by Meier (2001) for ADAF like accretion flows and later extended to the case of accretion disc coronae by Merloni and Fabian (2002).  It goes as follow, since  models and simulations of jet production (Blandford \& Znajek 1977;
Blandford \& Payne 1982, Ferreira 1997)  indicate that jets are driven 
by the poloidal component of the magnetic field. If we assume that the magnetic field is generated by dynamo processes in the disc/corona, the strengh of the poloidal component is limited by the scale height of the flow (Livio, Ogilvie \& Pringle 1999; Meier 2001; Merloni \& Fabian 2002). 
Therefore geometrically thick accretion flow should be naturally more efficient at launching jets.

Despite this strong link beween the corona and outflow in the LHS,
there are indications that in Intermediate States (IS) the jet  is connected to the accretion disc rather than the corona. For instance Malzac et al. (2006) report the results of an observation of Cygnus X-1 during 
a mini state transition. The dtat indicate that the  jet power is  anti-correlated with the disc luminosity and unrelated to the coronal power. This is in sharp contrast with previous results
obtained for the LHS, and suggests a different mode of coupling between the jet, the cold disc, and the corona in Intermediate States.

\section{Evolution of the geometry during outburst: the hysteresis problem}

Transient black hole binaries are a class of X-ray binaries that are detected  only occasionally when  they show some period of intense activity (outbursts) lasting for a few month (Tanaka \& Levin 1995). These outbursts have  recurrence times ranging from years to dozens of years. Between two episodes of activity the source is in an extremely faint quiescent state. During an outburst, the source luminosity  varies by many orders of magnitude. The study of spectral evolution of the sources during the outbursts is perfectly suited to understand how the structure of the corona depends on the mass accretion rate. The spectral evolution during an outburst is conveniently described using hardness intensity diagrams. Those diagrams show the evolution of the source luminosity as a function of spectral hardness (defined as the ratio of the observed count rates in a high to a lower energy band).
 Typically, during outbursts the sources follow a Q-shaped path in the hardness intensity diagram (see Figure~\ref{fig:hyste}). Initially in quiescence with a very low luminosity, they are in the LHS with a high hardness.  In the standard scenario\footnote{We note that a different geometrical interpretation of the outburts cycle in terms of a jet dominated accretion flow was recently proposed by Ferreira et al. (2006).
}, the cold disc is truncated very far away from the black hole, and the emission is dominated by the hot accretion flow. Then, as luminosity increases by several orders of magnitude the hardness remains constant.  
 The structure of the accretion flow is stable, the  cold disc inner radius probably moving slowly inward.  During this rise in X-ray luminosity  a compact jet is present and its radio emission  correlates with the X-ray emission. At a luminosity that is always above a few percent of Eddington,  the inner radius of the cold accretion disk decreases quickly.
 This reduction of the inner disc radius is associated with either
  the cold disk penetrating inside the hot inner flow, or the  later collapsing into an optically thick accretion disk  with small active regions of hot plasma on top of it 
   (Zdziarski et al. 2002). In both cases the enhanced soft photon flux from the disk 
   tends to cool down the hot phase, leading to softer spectra. The source therefore move from the right toward  the left hand side at almost constant luminosity.  During this hard to soft  state transition  the sources usually show a strong coronal activity (both the disc and corona are strong), this is also during this transition that optically thin radio flares associated with relativistic and sometimes superluminal ejections are observed. This region of the hardness intensity diagram is often called the Very High State (VHS). 
 Once a source has reached the upper left hand corner of the hardness intensity diagram, 
 it  never gets back to the LHS following the same path. Instead when the luminosity decreases, it stays in the HSS and when luminosity decreases, goes down vertically  and the strength of the non-thermal corona decreases, leaving a bare multiblackbody disc spectrum.
 This goes on until the luminosity reaches 0.02 percent of Eddington. Then it moves horizontally  to the right,  back to the LHS, and then down vertically back to quiescence.

It is to be noted that the luminosity of state transition from soft to hard seems to be relatively fixed around 
0.02 percent of Eddington (Maccaronne et al. 2003). On the the other hand the hard to soft transition can vary, even for the same source, by several order of magnitude and  it is always higher than the soft to hard transition luminosity. 
The question of what controls the different transitions luminosity is still open. Indeed, such a complex behaviour is not expected a priori, since in all accretion models, there is  only one relevant externally imposed parameter: the mass acretion rate. It was not anticipated that for the same 
luminosity the accretion flow could be in different states. It was recently suggeted that this complex behaviour could be explained by the existence of a second parameter that could be the magnetic flux advected with the accreting material Spruit \& Uzdensky (2005).
But the most detailed explanations so far are based on the idea that this second parameter is nothing else than the history of the system. In other word the accretion flow would present an hysteresis
behaviour because going from LHS to HSS is not the same thing as going from HSS to LHS. Within this line of reasoning, the most promising idea so-far was proposed by Meyer-Hofmeister et al. (2005)
 who suggested that the hysteresis could be linked to a condensation/evaporation equilibrium in the acretion disc corona system. They assume an advective corona on top of a thin optically thick accretion disc and  allowing for mass exchange between the corona and the disc. Depending on the temperature and density in the corona and the disc, the material in the corona may condensate into the disc and the disc may evaporate. Meyer-Hofmeister et al. (2005)  have computed the resulting equilibrium between an accretion disc and a corona as a function of the distance to the black hole and mass accretion rate (see also  R{\'o}{\.z}a{\'n}ska \& Czerny, 2000; Mayer \& Pringle, 2006).  They have found that at large distance the conditions are such that an equilibrium solution can always be found. At  closer distances from the black hole, disc evaporation becomes more and more important and, under some circumstances there is a critical radius below which the disc fully evaporate and no thin disc is possible. Below this radius we are then left with a pure ADAF. The location of this transisition radius depends on the mass accretion rate. At low mass accretion rate the transition radius is at 10$^3$ to 10$^4$   Schwarschild radii and decreases slowly as the the accretion rate increases until the critical mass accretion of about a few percent of Eddington is reached. Then, the inner radius suddenly drops to the last stable orbit: this is the hard to soft  state transition. As long as the source is in the LHS the flux of soft cooling photons in the corona is weak, and,  as a consequence the corona is hot and evaporation dominates over condensation. Once the HSS is reached however, there is a disc down to the last stable orbit emitting a strong thermal soft emission that efficiently cools the corona, keeping the condensation process strong even when the mass accretion rate decreases. As a consequence the transition from the soft to LHS occurs at a luminosity that is  about one order of magnitude lower  then the hard to soft transition. 

Although this model so-far provides the best explanation of the hysteresis behaviour,  it is not perfect.  First, the presence of the jet and possibly impacting on the disc corona equilibrium are completely ignored. Moreover this model does not explain why the transition LHS to HSS can occurs at different luminosities (the transition luminosity  is quite constrained in this model) and a second independent parameter may still be required.

\bibliographystyle{aa}

\end{document}